\documentclass[fleqn,usenatbib,useAMS]{mnras}

\usepackage{newtxtext,newtxmath}
\usepackage[T1]{fontenc}
\usepackage{ae,aecompl}

\usepackage{graphicx}	% Including figure files
\usepackage{grffile}  % Better handling of figure file names
\usepackage{amsmath}	% Advanced maths commands
\usepackage{amssymb}	% Extra maths symbols
\usepackage[usenames,dvipsnames]{color} %Text color

\newcommand{\Cee}{C}
\newcommand{\CorrC}{\Cee}
\newcommand{\CorrFourier}{\Cee_{\text{Fourier}}}
\newcommand{\CorrHaar}{\Cee_{\text{Haar}}}

\newcommand{\Nside}{\ensuremath{N_{\rmn{side}}}}
\newcommand{\psiHaar}{\ensuremath{\psi_{\rmn{Haar}}}}
\newcommand{\rmin}{\ensuremath{r_{\rmn{min}}}}
\newcommand{\rmax}{\ensuremath{r_{\rmn{max}}}}
\newcommand{\Mmax}{\ensuremath{M_{\rmn{max}}}}

\newcommand*{\code}[1]{\textsc{#1}}
\newcommand{\healpix}{\code{healpix}}
\newcommand{\camb}{\code{CAMB}}

\newcommand*{\satellite}[1]{\textit{#1}}
\newcommand{\WMAP}{\satellite{WMAP}}
\newcommand{\Planck}{\satellite{Planck}}

\newcommand*{\Planckmap}[1]{\texttt{#1}}
\newcommand{\smica}{\Planckmap{SMICA}}

\newcommand{\LCDM}{\ensuremath{\Lambda}CDM}

\renewcommand*{\vec}[1]{\bmath{#1}}
\newcommand*{\unitvec}[1]{\vec{\hat{#1}}}

\newcommand{\Shalf}{S_{1/2}}

\newcommand{\iimag}{\rmn{i}}
\newcommand{\eexp}{\rmn{e}}
\newcommand{\dderiv}{\rmn{d}}

\title[Suppressing large-angle correlations]{
Exploring suppressed long-distance correlations as the cause of suppressed large-angle correlations
}

\author[C.J. Copi, J. Gurian, A. Kosowsky, G.D. Starkman, and H. Zhang]
{
Craig J. Copi,$^{1,2}$\thanks{E-mail: cjc5@case.edu}
James Gurian,$^{1}$
Arthur Kosowsky,$^{4}$\thanks{E-mail: kosowsky@pitt.edu}
Glenn D. Starkman,$^{1,2,3}$\thanks{E-mail: glenn.starkman@case.edu}
\newauthor
and
Hezi Zhang$^{4}$
\\
$^{1}$Department of Physics, Case Western Reserve University, Cleveland, OH 44106-7079, USA\\
$^{2}$CERCA/ISO, Case Western Reserve University, Cleveland, OH 44106-7079, USA\\
$^{3}$Department of Astronomy, Case Western Reserve University, Cleveland, OH 44106-7079, USA\\
$^{4}$Department of Physics and Astronomy, University of Pittsburgh, Pittsburgh, PA 15260, USA
}

\date{Accepted XXX. Received YYY; in original form ZZZ}

\pubyear{2018}

\begin{document}
\label{firstpage}
\pagerange{\pageref{firstpage}--\pageref{lastpage}}
\maketitle

\begin{abstract}
  The absence of large-angle correlations in the map of cosmic microwave background temperature fluctuations is among the well-established anomalies identified in full-sky and cut-sky maps over the past three decades.
  Suppressed large-angle correlations are rare statistical flukes in standard inflationary cosmological models.
  One natural explanation could be that the underlying primordial density perturbations lack correlations on large distance scales.
  To test this idea, we replace Fourier modes by a wavelet basis with compact spatial support.
  While the angular correlation function of perturbations can readily be suppressed, the observed monopole and dipole-subtracted correlation function is not generally suppressed.
  This suggests that suppression of large-angle temperature correlations requires a mechanism that has both real-space and harmonic-space effects.
\end{abstract}


%%%%%%%%%%%%%%%%%%%%%%%%%%%%%%%%%%%%%%%%%%%%%%%%%%

%%%%%%%%%%%%%%%%% BODY OF PAPER %%%%%%%%%%%%%%%%%%

\section{Introduction}

Maps of the cosmic microwave background (CMB) temperature are currently the most informative of any cosmological data set.
The temperature power spectrum, now measured nearly to cosmic variance precision down to scales of a few arcminutes, matches the standard Lambda cold dark matter (\LCDM) cosmological model remarkably well, and fixes the basic cosmological model to high precision \citep{Jungman1996,Calabrese2013,Planck-R2-I}.
The CMB sky represents, to good approximation, an underlying scalar gravitational potential that is the realization of a  Gaussian random field, consistent with  `initial' conditions in the early Universe such as might have been laid down in some vanilla inflationary epoch.
Small distortions due to gravitational lensing \citep{Das2011,ACT-lensing,SPT-lensing,Planck-R2-XV} and the Sunyaev-Zel'dovich effect \citep{ACT-SZ,SPT-SZ,Planck-R2-XXII} have also been detected and characterized at high statistical significance, whereas non-Gaussianities that might have been generated during the inflationary epoch have not been detected.

\begin{figure}
  \centering
  \includegraphics[width=3.5in]{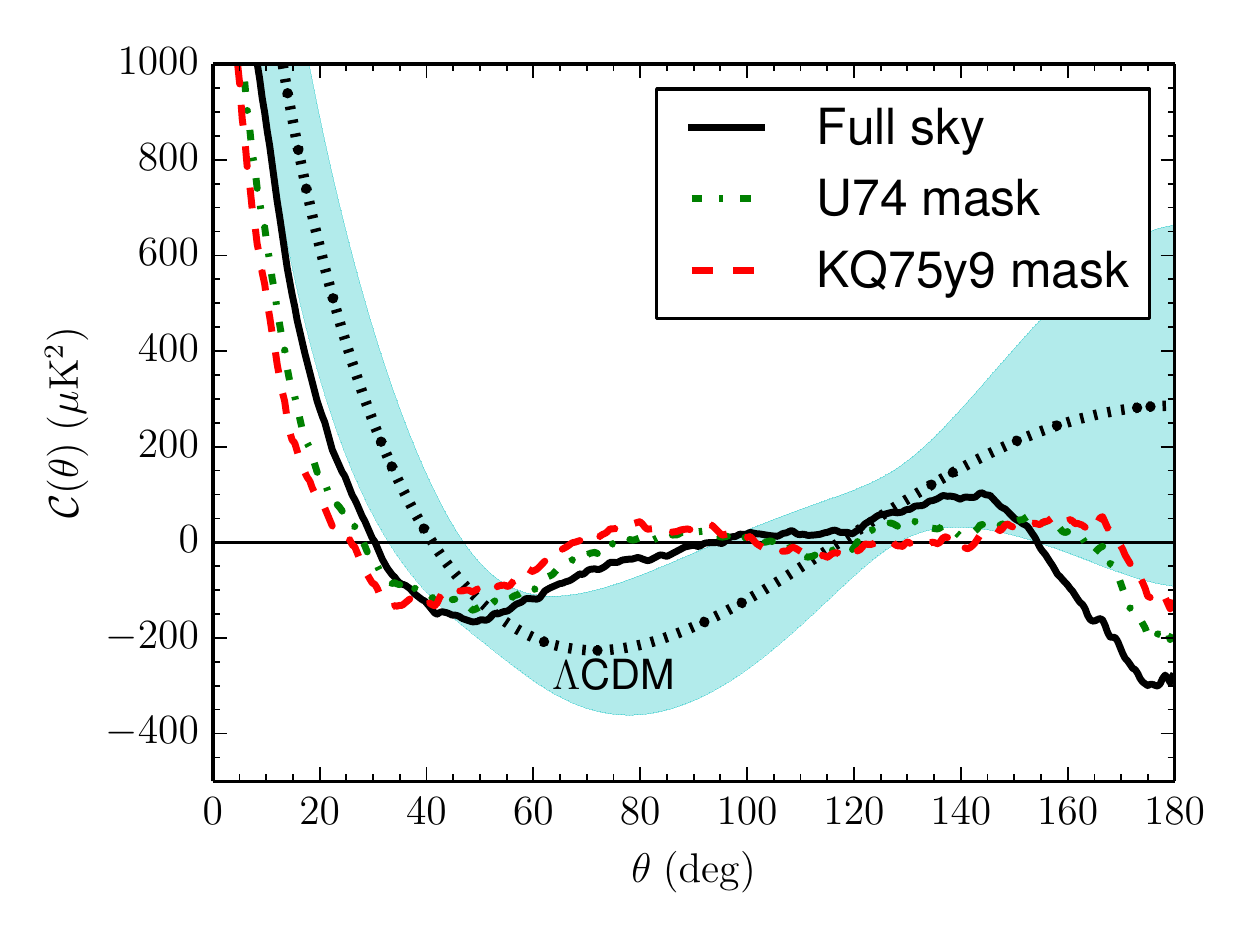}
   \caption{Two-point angular correlation function for the \Planck\ \smica\ map taken from \citealt{CHSS-Planck-R1-ctheta}.
    Shown is the best-fitting \LCDM\ model (dotted, black line) with its 68 per cent cosmic variance confidence interval (shaded, cyan region).
    The curves are the correlation function for the full-sky (solid, black line) and from two cut-skies: one using the \Planck\ U74 mask (dashed-dotted, green line) and the other using the \WMAP\ KQ75y9 mask (dashed, red line).
    Clearly seen is a lack of correlation on scales $\theta\ga 60\degr$.
  }
  \label{fig:Ctheta}
\end{figure}

Despite this broad success for inflationary cosmology, certain aspects of the microwave sky remain puzzling (see \citealt{Schwarz2016-Review} for a recent review).
Foremost among these may be the two-point temperature angular auto-correlation function
\begin{equation}
  \CorrC^{TT}(\theta) \equiv \left\langle T(\unitvec{e}_1) T(\unitvec{e}_2) \right\rangle,
\end{equation}
where $\langle\cdots\rangle$ represents an average over all pairs of directions on the sky, $\unitvec{e}_1$ and $\unitvec{e}_2$, having $\unitvec{e}_1 \cdot \unitvec{e}_2 = \cos\theta$.
In sky maps with the mean temperature and best-fitting dipole-subtracted, the angular correlation function is consistent with \textit{zero} for angles $\theta \ga 60\degr$, as shown in Fig.~\ref{fig:Ctheta} for the \Planck\ \smica\ sky map \citep{CHSS-Planck-R1-ctheta}.
This is quite puzzling, since random realizations of \LCDM\ with the observed power spectrum give correlation functions that are rarely so close to zero: a random sky in such a realization will have large-angle correlations  as small as those observed one less than once out of 1000 times \citep{CHSS-Planck-R1-ctheta} as quantified by reasonable statistics, such as
\begin{equation}
\Shalf\equiv \int_{-1}^{1/2}C(\theta)\, \dderiv(\cos\theta) ,
\end{equation}
first defined by the \WMAP\ team \citep{WMAP1-cosmology}.
Making the large-angle correlation function small while keeping the power spectrum approximately intact requires precise cancellations of the low-multipole contributions to the temperature angular power spectrum \citep{CHSS-WMAP5}.  However, these are uncorrelated in the standard \LCDM\ model.

This lack of large-angle temperature correlations may be a statistical fluke: an option that can be tested (see, e.g., \citealt{Dvorkin:2007jp,WMAP7-anomalies,CHSS-TQ,Yoho2014,Yoho2015,Copi2016-ISW,ODwyer2017-S4}).
However, its simple phenomenology suggests another possible explanation:
the primordial potential perturbations at the epoch of last scattering may themselves have zero expected correlations on scales larger than a given characteristic length.
The temperature anisotropies in the microwave background arise primarily from varying gravitational potentials at the epoch of last scattering (the Sachs-Wolfe effect), so the observed correlation would be significantly suppressed if the underlying perturbations lacked correlations.
The suppression would not be complete, due to contributions to the temperature anisotropies from line-of-sight effects, particularly the Integrated Sachs-Wolfe effect, but this is a subdominant contribution that does not spoil a purely Sachs-Wolfe explanation of the lack of large-angle correlations \citep{Copi2016-ISW}.

If the primordial correlation function is indeed suppressed on large length scales, the physics implications are profound.
This is not what is currently expected from early-universe inflation, which generically lasts much longer than required to solve the flatness and horizon problems.
It could be showing us the pre-inflationary gravitational-potential or the perturbations generated at the very beginning of inflation.
Alternately, it could signal that when inflationary perturbations are generated, they are coherent only over distances shorter than the causal horizon due to inflationary microphysics.
Either would be a fantastically valuable window into physical processes at energies far greater than we will likely ever obtain in colliders.
It would provide far richer information about the process of inflation (and perhaps how it began) than we can ever obtain if we can only probe inflation well after it is underway.

In this work we will explore the premise of this idea by way of a simple toy model.
We begin by observing that inflationary perturbations are conventionally characterized as the superposition of Fourier modes with amplitudes that are Gaussian-random, statistically independent, zero-mean variables.
However, each Fourier mode incorporates an infinite-range correlation.
We therefore replace the Fourier-mode basis by a basis of compact wavelets
with a maximum support.
We take the `cosmological' perturbations in this toy universe to be the superposition of wavelets with amplitudes that are Gaussian-random, statistically independent, zero-mean variables.
We will find that the expectation value of the correlation function vanishes trivially for separations exceeding the maximum support.
However, we will uncover a crucial complication -- the correlation function that vanishes at large angles is \emph{not} the monopole-and-dipole-subtracted correlation function, but the full correlation function.
Suppressing the large-angle monopole-and-dipole-subtracted angular correlation function requires suppressing the monopole and dipole fluctuations in the temperature power spectrum (i.e. $C_0$ and $C_1$).

This in no way follows automatically from the compact support of the basis functions, even in the mean.
Our results show that achieving this suppression of $C_0$ and $C_1$, without simultaneously suppressing the angular power spectrum at nearby $\ell$, while less improbable than in \LCDM, still seems not to be generic.
This conclusion does not appear to depend on the choice of wavelet family.
Our toy model's failings (eg. lack of statistical homogeneity) could be
responsible, although how so is not immediately obvious.
We identify these failings below, but reserve further exploration to future work.

Introducing a basis with compact support enforces a `spatial' condition on the fluctuations; also suppressing the monopole and dipole contributions to the CMB temperature fluctuations requires a `harmonic' condition.
In and of itself, this does not rule out suppression of large-distance correlations as the primary origin of the lack of large-angle correlations, but it requires physics that imposes constraints in both real and harmonic space.

\section{Toy Model: Wavelets in One Dimension}
\label{sec:toymodel-1d}

Wavelets provide a natural basis for expanding a function with compact support.
Here we will use the fact that wavelets have compact support in real space to remove correlations on large scales.
The simplest wavelet basis is the Haar wavelets.

Consider a function on the unit interval with periodic boundary conditions.
For this simple toy model, we can completely calculate the power spectra and correlation functions for both a Fourier and wavelet bases.
The usual Fourier expansion is
\begin{equation}
  \Phi(x) = \sum_{m=-\infty}^{\infty} \Phi_m \eexp^{2\upi\iimag m x}.
\end{equation}
For a real function $\Phi_{-m} = \Phi_{m}^*$, ignoring the monopole contribution ($m=0$) and imposing a maximum frequency (through a choice of $\Mmax$) we may rewrite this as
\begin{equation}
  \Phi(x) = 2\sum_{m=1}^{\Mmax} \rmn{Re}\left( \Phi_m \eexp^{2\upi\iimag m x}\right).
\end{equation}
Statistical independence in the Fourier basis corresponds to
\begin{equation}
  \langle{\Phi_m^* \Phi_{m'}} \rangle = C_m \delta_{m m'}.
\end{equation}
Throughout we will mimic a scale-invariant power spectrum, $C_m=1/m$.

\subsection{Haar Wavelets}

Instead of Fourier modes, we can alternatively expand a function in wavelets.
For simplicity, we use Haar wavelets, as they are the simplest wavelet basis and allow for analytic calculations to compare with numeric results.
The Haar mother wavelet is defined by
\begin{equation}
  \psiHaar(x) =
  \begin{cases}
    \hphantom{-}1, & 0 \le x < 1/2 \\
    -1, & 1/2 \le x < 1 \\
    \hphantom{-}0, & \mbox{otherwise}
  \end{cases} .
  \label{eq:Haar-mother}
\end{equation}
Any function may be expanded as
\begin{equation}
  \Phi(x) = \Phi_{-1} \psi_{-1}(x) + \sum_{r=\rmin}^{\rmax} \sum_{n=0}^{2^r-1} \Phi_{rn} \psi_{rn}(x),
  \label{eq:Phi-Haar-expansion}
\end{equation}
where $\Phi_{rn}$ are the expansion coefficients, $\psi_{-1}(x)=1$, and
\begin{equation}
  \psi_{rn}(x) = 2^{r/2} \psiHaar(2^r x - n).
\end{equation}
The set of functions $\{\psi_{rn}(x)\}$ form an orthonormal basis on the unit interval, $x\in[0,1).$
By construction, the periodic boundary conditions enforce $n\in[0,2^r-1]$.
The choice of $\rmin$ determines the maximum scale over which correlations can occur.
Statistical independence in the wavelet basis means that in the ensemble
\begin{equation}
  \langle \Phi_{rn} \Phi_{r' n'} \rangle = C_r \delta_{r r'} \delta_{n n'},
\end{equation}
where $C_r$ is the power spectrum of the fluctuations.
Though simple in form, the discontinuities in the Haar wavelet make them a poor choice for use beyond toy models.

For this toy example we will remove the monopole ($\Phi_{-1}=0$) and choose $\rmin=2$ corresponding to the suppression of correlations on scales larger than one quarter the circumference of the circle.
Further, to reduce computation time, we choose $\rmax=11$ which sets the sampling rate and thus the maximum frequency: $\Delta x = 2^{-\rmax-1}$ and $\Mmax = 2^{\rmax-1}$.

In one dimension, it is easy to study the dependence on the choice of wavelet basis.
The Daubechies wavelets provide such a convenient family.
The bases in this family are labelled by the highest number of vanishing moment -- with all moments up to the $n$th vanishing in db$n$.
In this notation, the Haar wavelets can be referred to as db1.
For comparison, we will include results from db2, a wavelet basis that is continuous, but not everywhere-differentiable.
Higher-$n$ db$n$ have also been explored and their behaviours are consistent with those found for db2.

\subsection{Power Spectrum Comparison}

As a first check, we verify that the power spectrum for Gaussian-random, statistically independent realizations in Fourier space, $C_m=1/m$, can be produced by choosing Gaussian random, statistically independent realizations in the wavelet basis following some power spectrum, $C_r$.
We find that this can be accomplished using
\begin{equation}
  C_r = \left( \frac{3}{2} \right) 2^{-r},
  \label{eq:Cr}
\end{equation}
as shown below.

By direct calculation it is found that, for $\Phi(x)$ expanded in the Haar basis, the Fourier coefficients are given by
\begin{align}
  \label{eq:Phim-Haar}
  \Phi_m \equiv&
  \int_{0}^{1} \Phi(x) \eexp^{-2\upi\iimag m x} \, \dderiv x \\
  =& \frac{1}{2\upi \iimag m} \sum_{r=\rmin}^{\rmax} \sum_{n=0}^{2^r-1} 2^{r/2} \Phi_{rn} \eexp^{-2\upi\iimag 2^{-r} mn} \left( 1 - \eexp^{-\upi \iimag 2^{-r} m}\right)^2. \nonumber
\end{align}
From this we find the power spectrum to be
\begin{equation}
  C_m \equiv \langle |\Phi_m|^2 \rangle
  = \frac{1}{(\upi m)^2} \sum_{r=\rmin}^{\rmax} 2^{2r+2} C_r \sin^4(2^{-r-1}\upi m).
  \label{eq:Cm-Haar}
\end{equation}

Alternatively, this can be tested numerically using a discrete wavelet transform and a fast Fourier transform.
The numerical test is of particular value for moving to other wavelet families where analytic calculations are not feasible.
In this case we have generated the $1000$ realizations of the wavelet coefficients according to the power spectrum~(\ref{eq:Cr}) and constructed $\Phi(x)$ using the Haar expansion~(\ref{eq:Phi-Haar-expansion}).
Recall that we have ignored the monopole ($\Phi_{-1}=0$) and are enforcing the lack of correlations on large scales by setting $\Phi_{rn}=0$ for $r<\rmin$.

\begin{figure}
  \centering
  \includegraphics[width=3.5in]{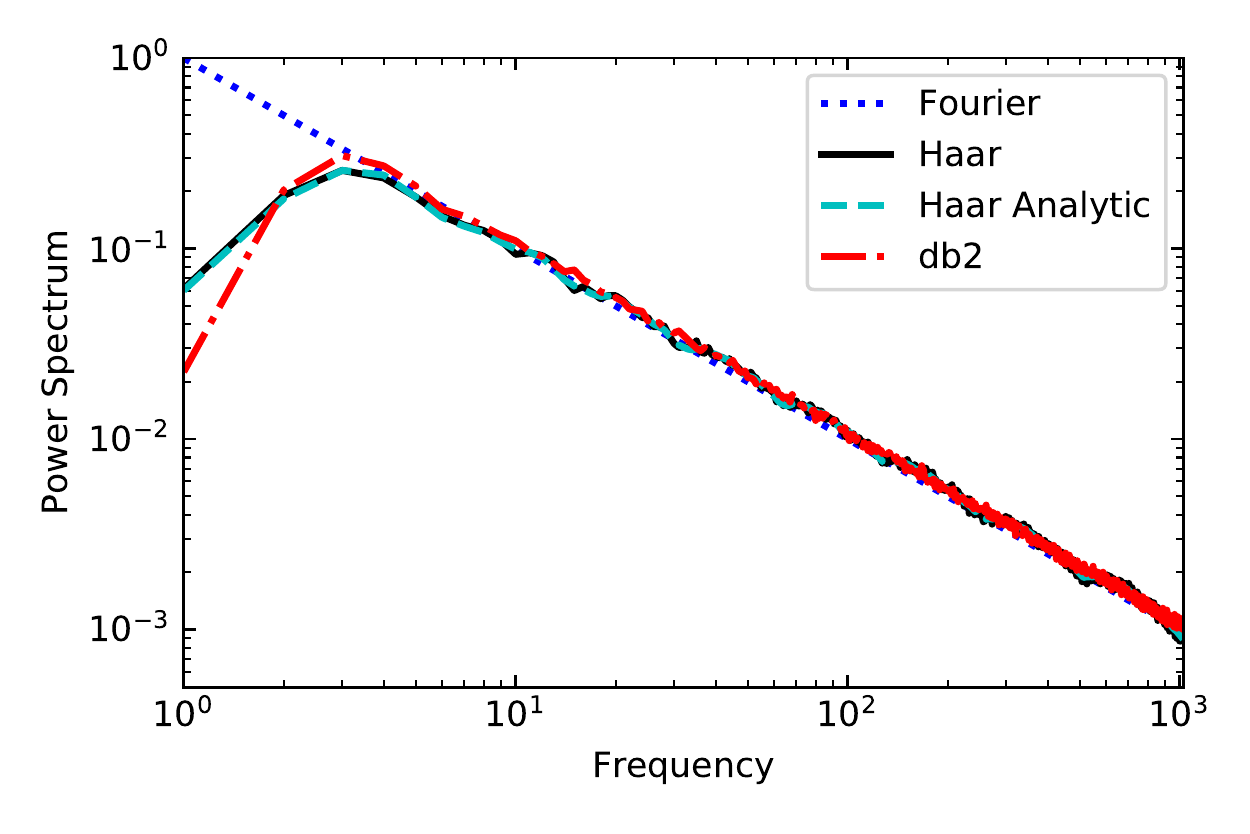}
  \caption{Power spectra in the Fourier, Haar, and db2 wavelet bases.
  The dotted, blue line represents the desired Fourier ($1/m$) behaviour.
  The solid, black line is the average of the $1000$ realizations in the Haar basis calculated using discrete transforms.
  Similarly, the dashed-dotted, red line is the average of $1000$ realizations in the db2 basis.
  Finally, the dashed, cyan line represents the analytic calculation~(\ref{eq:Cm-Haar}) using the power spectrum~(\ref{eq:Cr}) in the Haar basis.
  From the analytic calculation we see there is good agreement with the expected distribution over most of the frequency range.
  In all cases, the power spectra from the wavelet bases are in good agreement with the scale-invariant, Fourier power spectrum at all but the lowest frequencies.
  }
  \label{fig:power-spectra_1d}
\end{figure}
The results for the reconstructed power spectra are shown in Fig.~\ref{fig:power-spectra_1d}.
This figure shows the desired Fourier ($1/m$) power spectrum (dotted, blue line), the average from the realizations calculated using discrete Haar transforms (solid, black line), the result of the analytic power spectrum~(\ref{eq:Phim-Haar}) (dashed, cyan line), and the average from the realizations calculated using discrete db2 transforms (dashed-dotted, red line).
From this figure we see the analytic calculation closely follows the desired power spectrum over most the range, mainly deviating at lowest frequencies.
This deviation is mainly a suppression in the $m=1$ and $m=2$ modes, but not in any of the higher ones.
This is quite similar to what is found in the CMB where $C_2$ is small and a primordial $C_1$ is predicted to be small if the lack of correlation on large scales is physical \citep{CHSS-Planck-R1-ctheta}.
Overall we see that we \emph{can reproduce the power spectrum} with uncorrelated Gaussian random coefficients in a wavelet basis.

As further motivation for why changing the basis could produce the desired results of suppressing large-angle correlations without suppressing low-$\ell$ power, we note that the Fourier modes will now be correlated.
Explicitly, in the uncorrelated Gaussian-random wavelet basis, the correlations between different Fourier modes are now nonzero and given by
\begin{align}
  C_{m m'} \equiv& \langle \Phi_m^* \Phi_{m'} \rangle \\
  = & \frac{4}{\upi^2 m m'} \sum_{r=\rmin}^{\rmax} 2^r C_r \eexp^{\upi \iimag 2^{-r}(m-m')} \sin^2\!\left(\frac{m \upi}{2^{r+1}}\right) \sin^2\!\left(\frac{m' \upi}{2^{r+1}}\right)
  \nonumber \\
  & \qquad \times
  \begin{cases}
    2^r, & 2^{-r-1}(m-m')\in \mathbb{Z} \\
    \frac{1-\exp[\upi\iimag(m-m')]}{1-\exp[\upi\iimag 2^{-r}(m-m')]}, & \rmn{otherwise}
  \end{cases}.
  \nonumber
\end{align}
In principle, this effect could provide a natural explanation for the observed correlations between low-order multipole coefficients in the microwave background temperature maps \citep{CHSS-WMAP5}.

\subsection{Correlation Function}

\begin{figure}
  \centering
  \includegraphics[width=3.5in]{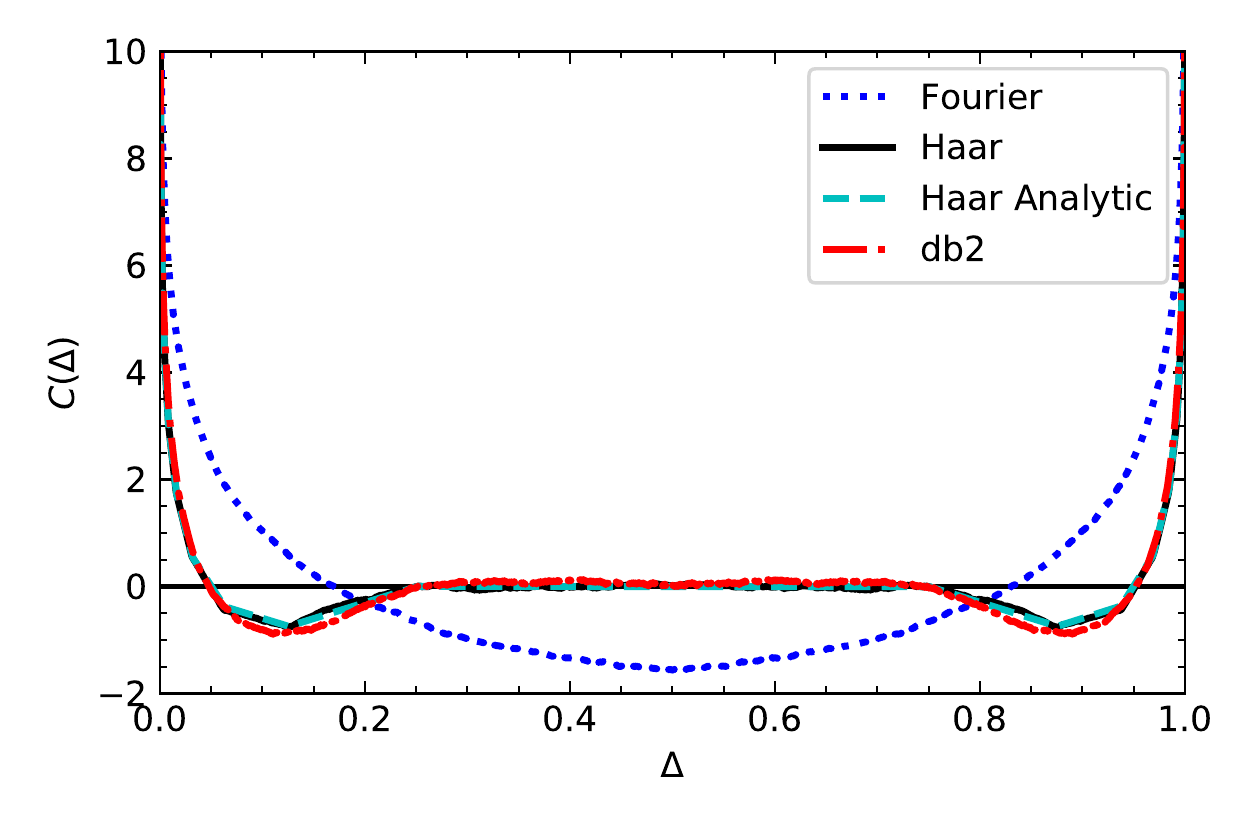}
  \caption{Correlation function for expansions in the Fourier and wavelet bases.
  Line styles and colours are the same as those in Fig.~\ref{fig:power-spectra_1d}
  The correlation function for the Haar basis is zero for $\Delta>1/4$, by construction, and is generally closer to zero over most of the range in $\Delta$ as compared to the that for Fourier basis.
  The correlation function from the db2 basis is nearly identical to that from the Haar basis.
  }
  \label{fig:CDelta}
\end{figure}

Our main interest is in the suppression of the correlation function on large scales.
We define the correlation-function estimator as
\begin{equation}
  \tilde \CorrC(\Delta) \equiv \int_{0}^{1} \Phi(x) \Phi(x+\Delta) \, \dderiv x.
\end{equation}
This estimator is unbiased, so its ensemble average equals the correlation function,
\begin{equation}
  \CorrC(\Delta) \equiv \langle \tilde \CorrC(\Delta) \rangle.
\end{equation}
For functions defined by uncorrelated Gaussian-random Fourier coefficients, the correlation function is
\begin{equation}
  \CorrFourier(\Delta) = 2 \sum_{m=1}^{\Mmax} C_m \cos(2\upi m \Delta).
\end{equation}
Similarly, for functions defined by uncorrelated Gaussian-random Haar wavelet coefficients, the correlation function can be written as
\begin{equation}
  \CorrHaar(\Delta) = \sum_{r=\rmin}^{\rmax} 2^r C_r \times
  \begin{cases}
    \hphantom{-}1 - 3(2^r \Delta), & 0 \le 2^r \Delta < 1/2 \\
    -1 + 2^r \Delta, & 1/2 \le \Delta < 1 \\
    0, & \rmn{otherwise}
  \end{cases}.
\end{equation}
From Fig.~\ref{fig:CDelta}, we see that in the Haar basis $\CorrHaar(\Delta) = 0$ for $1/4\le\Delta\le1/2$, as required by construction.
The periodic boundary condition introduces a reflection symmetry across the line $\Delta=1/2$.
Overall, $\CorrHaar(\Delta)$ is nearer to zero for most the of the range of $\Delta$, as compared to the correlation function in the Fourier basis.
This is necessary but not sufficient to ensure that a typical realization has suppressed correlation on large scales.

\section{Toy Model: Wavelets in Three Dimensions}
\label{sec:toymodel-3d}

The results from the one-dimensional study are encouraging.
We see that we can reproduce a scale-invariant power spectrum while suppressing correlations on large scales.
Unfortunately the one-dimensional case is not sufficient to model the CMB\@.
Encouraged by these results, we now move to three dimensions.
We generate a field in three dimensions that lacks correlations on large length scales.
To mimic the CMB, we consider the statistical properties of the field on the surface of a sphere.
This mimics the last scattering surface (LSS), and thus is modelling the Sachs-Wolfe~(SW) contribution.
We will ignore the integrated Sachs-Wolfe~(ISW) contributions.
Provided that the power spectrum at small scales (large $\ell$) are consistent with \LCDM, the ISW will also be consistent with \LCDM\@.

\subsection{Haar Wavelets}

Again ignoring the monopole, we expand a field in three dimensions as the product of three one-dimensional wavelets,
\begin{equation}
  \Phi(\vec x) = \sum_{\vec r=\rmin}^{\rmax} \sum_{\vec n} \Phi_{\vec r \vec n} \psi_{r_x n_x} (x) \psi_{r_y n_y}(y) \psi_{r_z n_z}(z).
\end{equation}
Here we use shorthand notation of $\vec r\equiv\{r_x, r_y, r_z\}$ and similarly for $\vec n$.
We again assume statistical independence in the Haar basis so that
\begin{equation}
  \langle \Phi_{\vec r \vec n} \Phi_{\vec r' \vec n'} \rangle = C_r \delta_{\vec r \vec r'} \delta_{\vec n \vec n'}.
\end{equation}
As in the one-dimensional case we will choose
\begin{equation}
  C_r = \left( \frac{A}{\sqrt{2^{2r_x} + 2^{2r_y} + 2^{2r_z}}} \right)^3
\end{equation}
to mimic the scale independent, three-dimensional Fourier power spectrum, $P(k)\propto k^{-3}$.
Note that the normalization, $A$, is arbitrary and will be chosen to match the observed CMB power spectrum.

Realizations of the field, $\Phi(\vec x)$, can now be constructed at all points in space.
As mentioned above, a key difference between the one-dimensional and three-dimensional cases is that the field is to be evaluated on a sphere.
We again choose a unit interval, in this case a unit box, in which to calculate field.
The sphere representing the LSS is chosen to have a radius $R<1/2$ to fit completely inside the box (alleviating the need for specifying boundary conditions at the box edges) and is located at
\begin{equation}
  \vec x_c = \left( \frac{1}{2}, \frac{1}{2}, \frac{1}{2} \right) + \Delta \vec x_c.
\end{equation}
In other words, $\Delta \vec x_c=0$ when the centre of the sphere is at the centre of the box.

We thus have two extra parameters to consider when generating realizations: $R$ and $\Delta x_c$.
Note that there is a degeneracy between changing $\rmin$ and changing $R$.
To be consistent with the one-dimensional case we will continue to use $\rmin=2$ and only consider changes in $R$.

The procedure for generating realizations in three dimensions is now conceptually straightforward.
Choose a radius for the sphere, $R$, a location for the centre of the sphere, $\Delta\vec x_c$, a minimum Haar resolution, $\rmin$, a maximum Haar resolution, $\rmax$, and the \healpix\footnote{See \url{http://healpix.sourceforge.net} resolution on the sphere, $\Nside$.
With these choices the field $\Phi(\vec x)$ is evaluated at the centre of \healpix\ for more information.} pixels.
In practice, the challenge to generating realizations comes from the number of Haar coefficients required for describing $\Phi(\vec x)$ in the box.
However, by only generating the coefficients, $\Phi_{\vec r \vec n}$, needed for the pixels on the sphere, and only generating and storing them while they are required, greatly reduces the time and memory required.
Once the procedure is completed we are left with a scalar map on the sphere.
This represents the Sachs-Wolfe contribution to the CMB and can be analysed using the standard tools provided by \healpix.

\subsection{Results}

We will consider two different choices for the radius of the sphere, $R$, and choose the rest of the parameters as follows: $\Delta \vec x_c=0$, $\rmin=2$, $\rmax=10$, and $\Nside=512$.
Unlike in the one-dimensional case, in the three-dimensional case we can normalize the results to those from $\LCDM$ to allow direct comparisons in known units.
Since the Haar power spectrum reproduces a scale-invariant power spectrum in the mean at small scales, we can normalize the small scale behaviour of the Haar power spectrum to the small scale behaviour of the best-fitting \LCDM\ power spectrum.
To find the normalization constant, $A$, we require the average of the power spectrum,
\begin{equation}
  D_\ell \equiv \frac{\ell(\ell+1)}{2\upi} C_\ell,
\end{equation}
over the range $55\le\ell\le 65$ to be one.
This range is sufficiently high as to avoid differences at low-$\ell$ and sufficiently low as to avoid finite resolution effects at high-$\ell$.
To include the transfer function we multiply this normalized power spectrum by the best-fitting \LCDM\ power spectrum calculated from the \camb\footnote{See \url{http://camb.info/} for information
  and access to the code.} code \citep{2000ApJ...538..473L} without inclusion of the late-time ISW\@.
In practice, the true Haar power spectrum should be evolved using a Boltzmann code in place of the (nearly) scale-invariant one employed here.
However, this normalization is only meant to help guide the eye when qualitatively comparing results.

\begin{figure*}
  \centering
  \hbox{
  \includegraphics[width=3.5in]{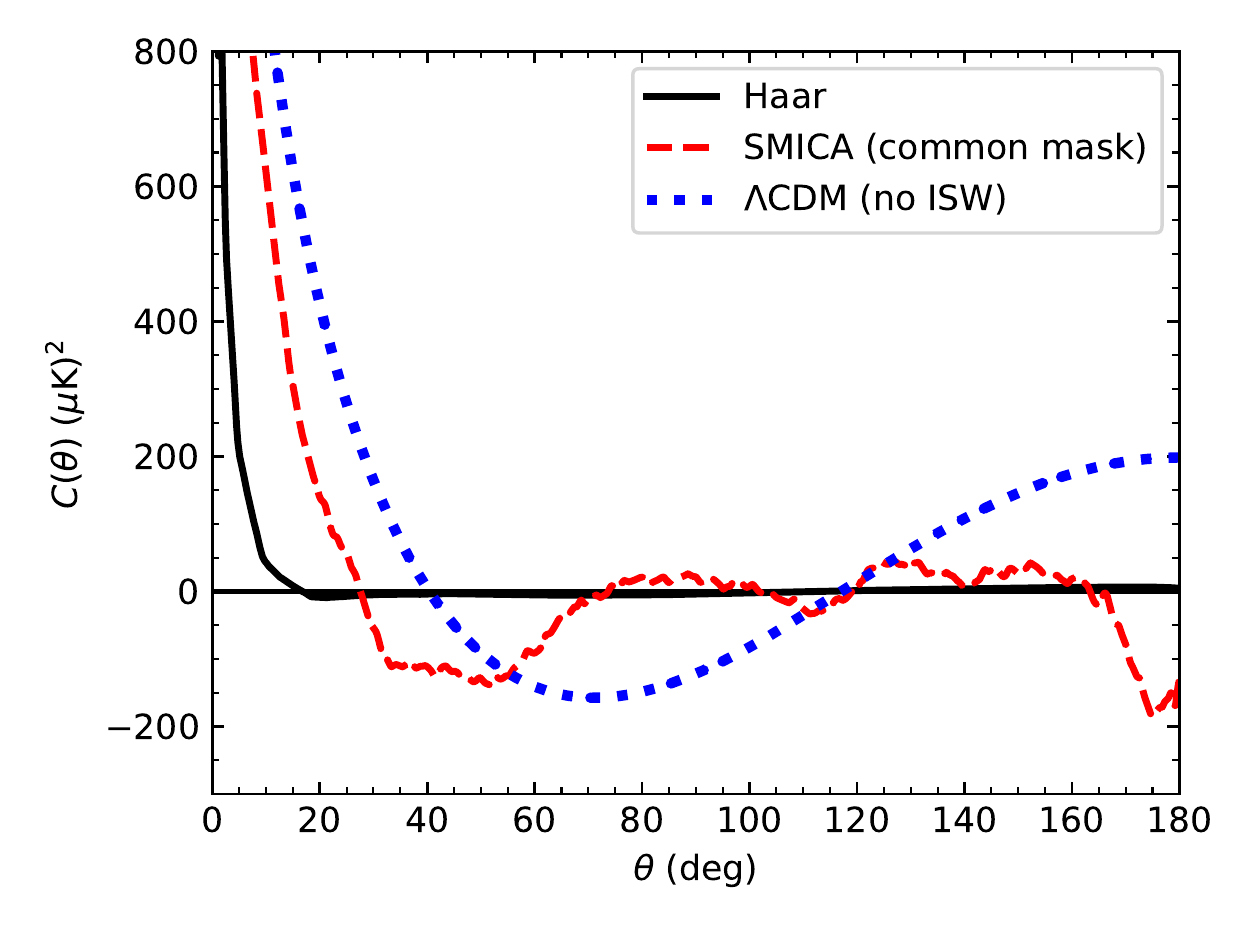}
  \hfill
  \includegraphics[width=3.5in]{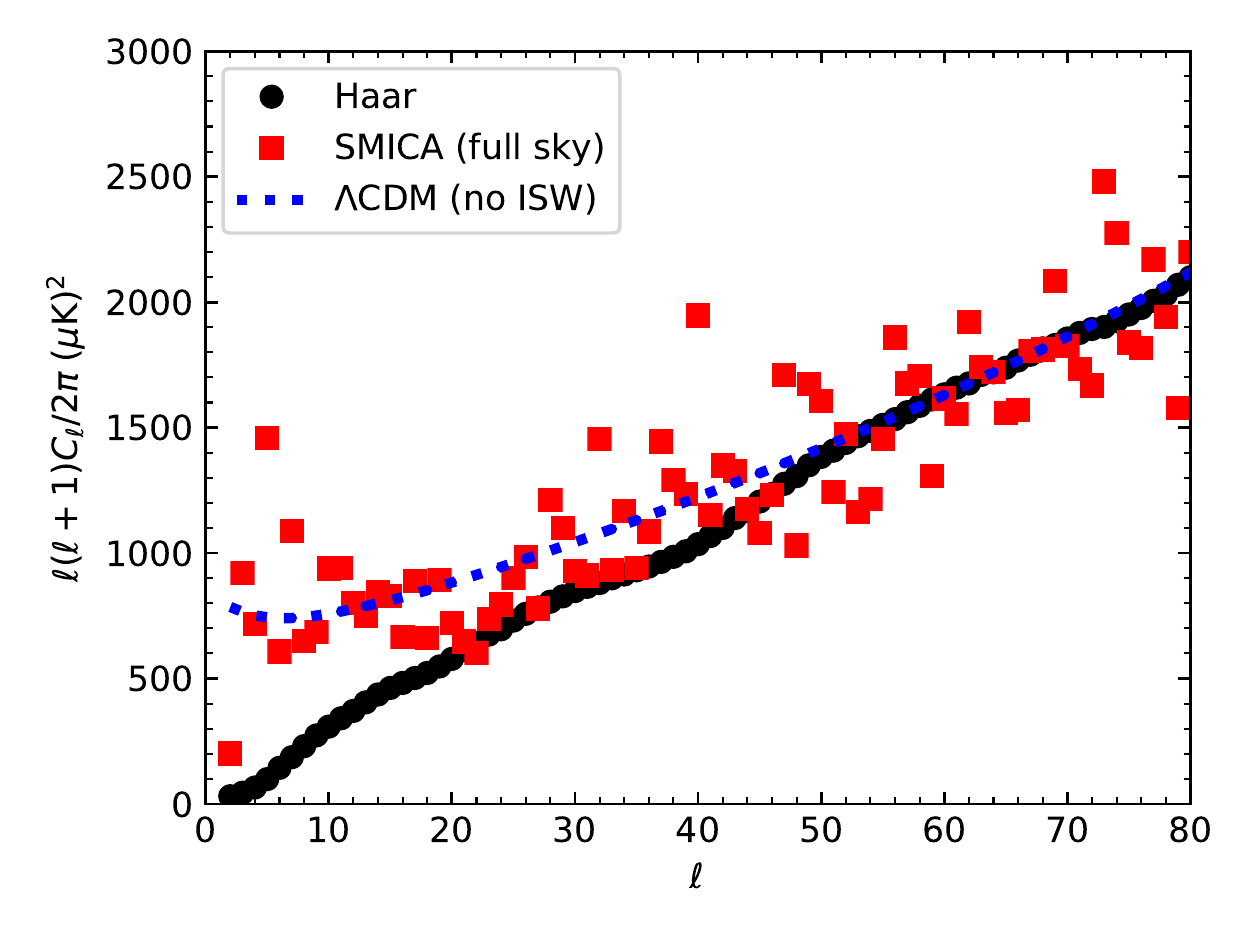}
 }
  \caption{Results from 5000 realizations of Haar wavelets in three dimensions for $\rmin=2$, $\rmax=10$, $\Delta\vec x_c=0$, $\Nside=512$, and a sphere of radius $R=0.2$.
  \emph{left panel:} Two-point angular correlation function normalized to \LCDM\ without ISW contributions (shown as the blue, dotted line).
  The Haar correlation function (black, solid line) is greatly suppressed, essentially zero for $\theta\ga 20\degr$.
  For reference the cut-sky \Planck\ R3 \smica\ correlation function is also shown (red, dashed line).
  \emph{right panel:} Two-point angular power spectrum normalized to \LCDM\ without ISW contributions (shown as the blue, dotted line).
  The Haar values (black circles) are suppressed at low-$\ell$ compared to \LCDM\@.
  For reference to data, the full-sky \Planck\ R3 \smica\ power spectrum is also shown (red squares).
  Despite containing ISW contributions, it is in good agreement with the \LCDM\ curve.
  }
  \label{fig:Haar-3d-R0.2}
\end{figure*}

As a first case we consider a sphere with $R=0.2$.
The results from 5000 realizations are shown in Fig.~\ref{fig:Haar-3d-R0.2}.
The panel on the left shows the average of the two point angular correlation function for the Haar realizations (solid, black line) along with the cut-sky two point angular correlation function from the \Planck\ Release 3 \smica\ map using the common mask \citep{Planck-R3-IV}.
Note that the Haar realizations do not include the late-time ISW, unlike the data.
For qualitative comparisons this is fine.
The Haar two-point angular correlation function is almost completely suppressed above about $20\degr$, thus, including the late-time ISW would produce a correlation function that is completely dominated by the late-time ISW on large angular scales.
As noted in the introduction, it was shown by \citet{Copi2016-ISW} that suppression of the SW contribution is sufficient to suppress correlations consistent with the data as quantified by the $\Shalf$ statistic.
Thus we see that the two-point angular correlation function is suppressed on large angular scales.

The right panel of Fig.~\ref{fig:Haar-3d-R0.2} shows the two-point angular power spectra for the Haar realizations (black circles) and the full-sky \Planck\ Release 3 \smica\ map.
Again, the realizations are purely the SW contributions, whereas the data contains all contributions.
Here we see that the Haar power spectrum is greatly suppressed compared to the (nearly) scale-invariant expectation (as represented by the blue, dotted line) and also the data.
Thus the suppression of the large-angle, two-point angular correlation function has been achieved by suppressing the power spectrum at low-$\ell$\@.
This is not the desired behaviour, nor is it consistent with the data.

\begin{figure*}
  \centering
  \hbox{
  \includegraphics[width=3.5in]{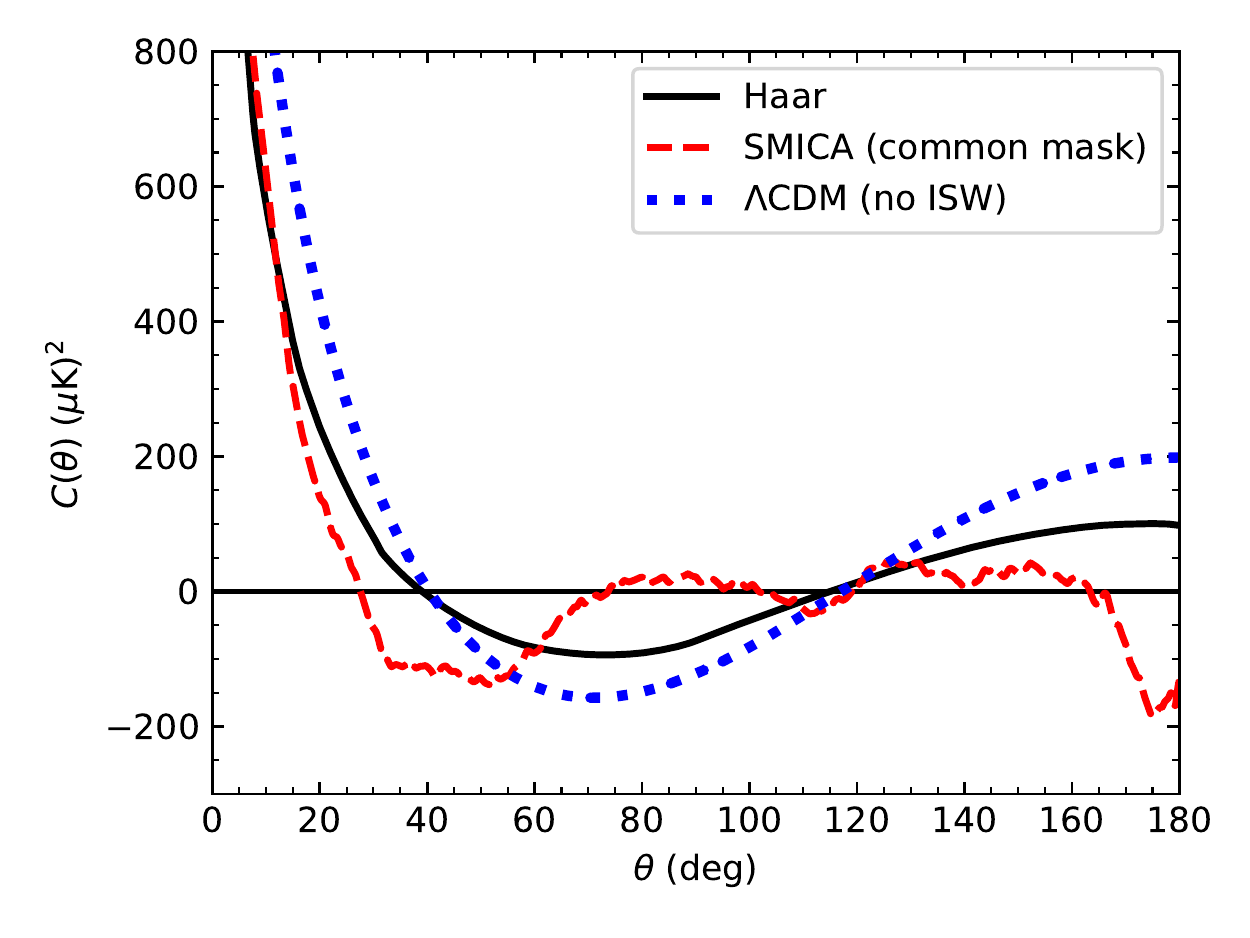}
  \hfill
  \includegraphics[width=3.5in]{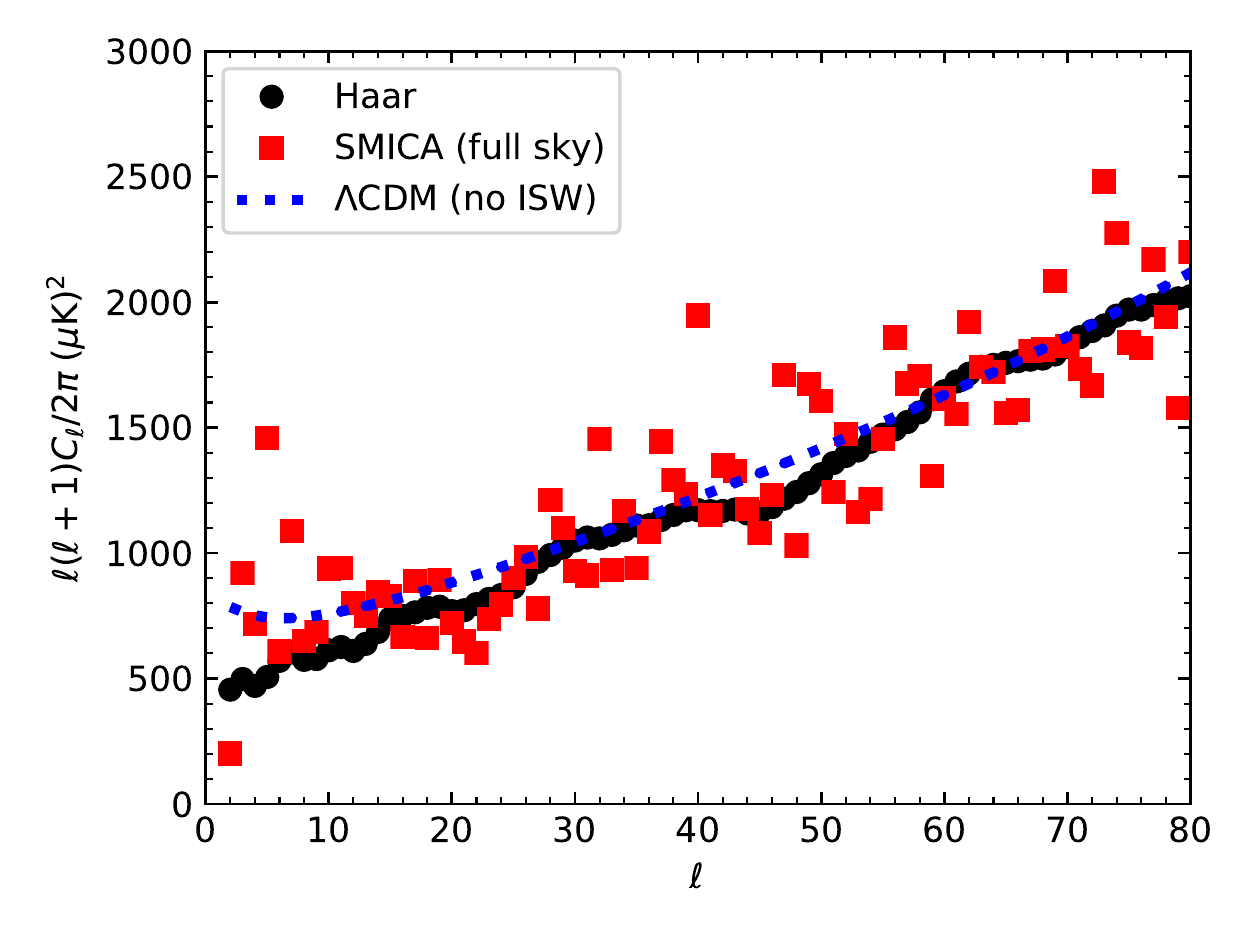}
  }
  \caption{Similar to Fig.~\ref{fig:Haar-3d-R0.2} but now for a sphere of radius $R=0.03$.
  For this case the two-point angular power spectrum (right panel) is now qualitatively in good agreement with the data, even at low-$\ell$\@.
  However, the two-point angular correlation function (left panel) is now much more similar to the \LCDM\ curve showing only a moderate suppression.
  At first glance it is surprising that the correlation function is not zero about $90\deg$!
  See text for a discussion.
  }
  \label{fig:Haar-3d-R0.03}
\end{figure*}

The power spectrum can be made more consistent by reducing the size of the sphere.
Fig.~\ref{fig:Haar-3d-R0.03} shows the results for a sphere of radius $R=0.03$\@.
The power spectrum (right panel) is now much more consistent with the data, though still somewhat suppressed compared to the (nearly) scale-invariant power spectrum.
Unfortunately, this comes at the cost of the correlation function (left panel) no longer being suppressed.
In fact, the correlation function looks similar in shape to that from \LCDM\@.
The correlations in the Fourier modes induced by the Haar basis are not sufficient to suppress the correlation function when the power spectrum is not suppressed.
In other words, the suppression of the correlation function in the Haar basis is achieved by suppressing power at low-$\ell$.

This behaviour persists when the parameter space is more thoroughly explored.
Choosing different values for the radius, $R$, and the location of the centre of the sphere, $\Delta\vec x_c$, does change the correlation function and the power spectrum, but in ways consistent with that shown above.
It remains true that the suppression of the large-angle correlation function always comes at the expense of suppressing low-$\ell$ power.

\section{Role of the Monopole and Dipole}

At first glance the fact that the two-point angular correlation function in Fig.~\ref{fig:Haar-3d-R0.03} (left panel) is not zero is surprising.
The original idea was to completely remove correlations on large scales by the choice of basis.
How can the correlation function be non-zero?
Even more strongly, locating the sphere at the centre of the box is a special choice.
For $\rmin=2$ the correlation function \emph{must} be zero in the mean for $\theta>90\degr$ by construction since there are no Haar modes that cross any of the planes $x=1/2$, $y=1/2$, or $z=1/2$.
Thus it is impossible to have correlations on scales larger than $90\degr$.
What happened?

The problem is that what we actually construct from the CMB data is the monopole and dipole-subtracted two-point angular correlation function.
The monopole is subtracted by necessity: we are actually calculating the correlation function of temperature fluctuations.
The dipole is subtracted since observationally it is dominated by our motion (the Doppler dipole).
The \emph{full} correlation function, the one that includes the monopole and dipole contributions, \emph{is zero} as it must be.
Unfortunately, that is not what we actually need to suppress to be consistent with the data.

The monopole fluctuation represents the difference between the average temperature of our CMB sky, and the the average temperature of the CMB sky of some other observer, including one who is not in causal contact with us.
In inflationary \LCDM, this monopole fluctuation would arise automatically from the contributions of Fourier modes over long distances.
At any given time, it cannot be observationally disentangled from the background average temperature.
The dipole fluctuation arises similarly in \LCDM, and cannot (at least currently) be accurately separated from the Doppler contribution to the temperature dipole \citep{2011PhRvL.106s1301K,2017PhRvL.119v1102Y}.
Even though they cannot be measured, the monopole and dipole temperature fluctuations still have effects.
The large-angle temperature-temperature correlation function and the large-angle monopole and dipole-subtracted TT correlation function can both be small if and only if the contributions of the monopole and dipole temperature fluctuations to the correlation function are also small.
This further shows the peculiarity of our Universe's small large-angle monopole and dipole-subtracted temperature-temperature correlation function.

\section{Conclusions}

The anomalous lack of large-angle correlations in the CMB temperature maps is suggestive of a lack of large-distance correlations in the underlying scalar gravitational potential.
In this work, we have explored this possibility by replacing the traditional harmonic bases for cosmological fluctuations with a selection of wavelet bases, in which each mode has compact support.
While Fourier modes inherently encode infinite distance correlations, and spherical harmonics encode correlations across the whole sky, the wavelet bases are partially localized in momentum space, but fully localized in real space.
By enforcing a hard cutoff on the size of the largest wavelet that was smaller than the diameter of the last scattering surface (LSS), we automatically ensured that, in the mean, the angular correlation function vanished on large scales and has a small variance.

However, crucially, our explorations revealed a key difference between the vanishing of the angular correlation function on large angular scales, and what is actually observed in data -- the vanishing of the angular correlation function of the monopole-and-dipole-subtracted CMB sky.
In order for these two to approximately coincide, the fluctuations in the monopole and dipole must themselves be small!
The suppression of these multipoles is, by definition, a condition on the harmonic components of the sky, and not automatically enforced by the lack of large spatial correlations in the wavelet basis.
By adjusting the position of the centre of the LSS relative to the `grid' of the wavelet basis, and by adjusting the size of the largest wavelet relative to the size of the LSS, it was possible to reduce the values of $C_0$ and $C_1$, however it was difficult to do so without simultaneously suppressing $C_2$, $C_3$, and other low-$\ell$ $C_\ell$.
While $C_2$ is indeed low in the observed CMB, and $C_3$ is at least low on the sky with the Galaxy masked \citep{WMAP1-PS}, other low-$\ell$ $C_\ell$ are not observed to be low.
It thus appears to remain difficult (albeit less so than in standard \LCDM) to obtain a large p-value for low $\Shalf$, the standard measure of large-angle correlations, while keeping the angular power spectrum consistent with observations.

The toy model we have explored has obvious shortcomings.
Most importantly, the wavelet basis explicitly violates statistical homogeneity -- it is constructed on a fixed grid -- and statistical isotropy -- the basis modes have a preferred orientation.
It is possible that a perturbation basis which more accurately reflects the underlying cosmological symmetries could improve the model, but the mechanism by which this might be accomplished is unclear.

We note that an alternative explanation of finite large-angle correlations that shows some promise is that the topology of the Universe is non-trivial.
Non-trivial topology alters the boundary conditions of the wave equation underlying the traditional basis choice of Fourier modes or spherical Bessel functions times spherical-harmonics, and constrains correlations on scales larger than the `topology scale'.
In \citet{Bernui}, it was shown that a slab topology (flat geometry with the topology of infinite Euclidean two-space crossed with a circle) with an identification scale of approximately $1.4$ times the radius of the last scattering surface would suppress large-angle correlations, while not inappropriately altering the low-$\ell$ angular power spectrum.
(A similar topology was also studied in \citealt{2004PhRvD..69f3516D}.)
However, as noted by the authors of \citet{Bernui}, that specific possibility runs afoul of existing circles-in-the-sky \citep{Cornish:1997hz,Cornish1998-circles} limits from \WMAP\ \citep{Cornish2004,ShapiroKey:2006hm,Vaudrevange} and \Planck\ \citep{Vaudrevangeprivate}.

Independent of a model of primordial perturbations which naturally produces the observed suppression of large-angle correlations, testing the fluke hypothesis with microwave background polarization and lensing data remains compelling \citep{CHSS-TQ,Yoho2014,Yoho2015}.

\section*{Acknowledgements}
GDS is partially supported by Department of Energy grant DE-SC0009946 to the particle astrophysics theory group at CWRU.
Some of the results in this paper have been derived using the \healpix\ \citep{2005ApJ...622..759G} package.
This work made use of the High Performance Computing Resource in the Core Facility for Advanced Research Computing at Case Western Reserve University.
The authors thank Amanda Yoho and Simone Aiola for discussions in a preliminary incarnation of this work.

%%%%%%%%%%%%%%%%%%%%%%%%%%%%%%%%%%%%%%%%%%%%%%%%%%



\bibliographystyle{mnras}
\bibliography{cmb_wavelets}

%%%%%%%%%%%%%%%%%%%%%%%%%%%%%%%%%%%%%%%%%%%%%%%%%%

%%%%%%%%%%%%%%%%% APPENDICES %%%%%%%%%%%%%%%%%%%%%
%\appendix

%%%%%%%%%%%%%%%%%%%%%%%%%%%%%%%%%%%%%%%%%%%%%%%%%%

% Don't change these lines
\bsp	% typesetting comment
\label{lastpage}
\end{document}